\documentclass{aa}

\newcounter{sub}
\newcounter{subeqn}[sub]
\def\be{\begin{equation}}
\def\ee{\end{equation}}
\def\bt{\begin{tabular}}
\def\et{\end{tabular}}
\def\lp{\left(}
\def\rp{\right)}
\def\ls{\left[}
\def\rs{\right]}

\def\st{\stepcounter{sub}}
\def\stq{\stepcounter{subeqn}}
\def\bea{\begin{eqnarray}}
\def\eea{\end{eqnarray}}
\def\beas{\begin{eqnarray*}}
\def\eeas{\end{eqnarray*}}

\def\x{{\bf x}}
\def\r{{\bf r}}

\def\v{{\bf v}}

\def\xp{{\bf x'}}

\def\Y{{\bf Y}}
\def\xxi{\xi\hspace{-.18cm}\xi\hspace{-.18cm}\xi}
\def\zzeta{\zeta\hspace{-.19cm}\zeta\hspace{-.19cm}\zeta}
\def\OOmega{\Omega\hspace{-.245cm}\Omega\hspace{-.245cm}\Omega}
\def\nab{\nabla\hspace{-.305cm}\nabla\hspace{-.305cm}\nabla}

\def\d{\delta}

\begin{document}

   \thesaurus{ 08     
              (
               08.14.1;  
               08.15.1;   
               08.18.1   
                       )}

   \title{The r-modes of rotating fluids}


   \author{Y. Sobouti
          \inst{1,2}
          \and
          V. Rezania
          \inst{1,3}
          }

   \offprints{V. Rezania \\ $^*$ Present address}

    \institute{Institute for Advanced Studies in Basic Sciences,
               Gava Zang, Zanjan 45195, Iran\\
          \and
              Center for Theoretical Physics and Mathematics,
               AEOI, P.O. Box 11345-8486, Tehran, Iran   \\
          \and
              Department of Physics, University of Alberta,
              Edmonton AB, Canada T6G 2J1 $^*$ \\
              emails: sobouti@iasbs.ac.ir \& vrezania@phys.ualberta.ca
             }

   \date{Received / accepted }

   \maketitle

   \begin{abstract}

An analysis of the toroidal modes of a rotating fluid, by means of
the differential equations of motion is not readily
tractable.  A matrix representation of the equations in a
suitable basis, however, simplifies the problem considerably and
reveals many of its intricacies.  Let $\Omega$ be the
angular velocity of the star and ($\ell,\,m$) be the two
integers that specify a spherical harmonic function.  One readily
finds the followings:

1) Because of the axial symmetry of equations of motion, all
modes, including the toroidal ones, are designated by a definite
azimuthal number $m$.

2) The analysis of equations of motion in the lowest order of
$\Omega$ shows that
Coriolis
forces turn the neutral toroidal
motions of ($\ell,\,m$) designation of the non rotating fluid into
a sequence of oscillatory modes with
frequencies $2m\Omega/\ell(\ell +1)$.
This much is a common knowledge.  One can say more, however.
a) Under the Coriolis forces,
the eigendisplacement vectors remain purely toroidal and carry the
identification ($\ell,m$).   They remain decoupled from other toroidal
or poloidal motions belonging to different $\ell$'s.
b) The eigenfrequencies quoted above are still degenerate.  For they carry no
reference to a radial wave number.
As a result the eigendisplacement vectors, as far as their radial dependencies
go, remain indeterminate.

3) The analysis of equation of motion in the next higher order
of $\Omega$ reveals that the forces
arising from asphericity of the fluid and the square of the
Coriolis terms (in some sense) remove the radial degeneracy.  The
eigenfrequencies now carry three identifications ($s,\ell,m$),
say, of which $s$ is a radial eigennumber.  The eigendisplacement
vectors become well determined.  They still remain zero order and purely
toroidal motions with a single ($\ell,m$) designation.

4) Two toroidal modes belonging to $\ell$ and $\ell\pm 2$ get coupled only at
$\Omega^2$ order.

5) A toroidal and a poloidal mode belonging to $\ell$ and $\ell\pm
1$, respectively, get coupled but again at $\Omega^2$ order.

Mass and mass-current multipole moments of the modes that are
responsible for the gravitational radiation, and bulk and shear
viscosities that tend to damp the modes are worked out in much
details.

   \end{abstract}

%

\section{Introduction}\label{int}

Recent years has witnessed a surge of interest in the small
oscillations of rotating fluid masses. The reason for the
excitement is the advocation by relativists that in rapidly
rotating neutron stars the gravitational radiation  drives the
$r$-modes unstable, and while spinning down the star, may itself
be amenable to detection
(see recent review by Andersson and Kokkotas \cite{AKo00}).
Nevertheless, the oscillations of
rotating objects is an old problem. In the past few decades it
has been studied by many investigators and from various points of
view. Complexity of the problem arises from the fact that a fluid
can support three distinct types of motions, derived from, say, a
scalar potential, from a toroidal vector potential and from a
poloidal vector potential. These are the motions associated
predominantly with the familiar $p$-, $g$- and toroidal-
oscillations of the fluid. Each of these motions, in turn, can be
given an expansion in terms of vector spherical harmonics. The
modes of an actual star are a mixture of the three types mentioned
above and of the different spherical harmonic components. Sorting
out this mixture and classifying the modes into well defined
sequences has not been an easy task. Moreover, and more often than
not, it has not been realized that $g$- modes of spherically
asymmetric configurations are not apt for perturbation analysis.
For, the low frequency tail of their spectrum
is a fragile structure. It is driven by minute buoyancy forces and
can be completely wiped out by almost any perturbing agent such as
Coriolis, asphericity and magnetic forces.

Here we show that a matrix representation of the equations of
motions provides a set of algebraic equations that are much easier
to cope with than their differential counterparts. In section 2 we
write down the equilibrium structure and the linearized equations
of motion of a rotating star. In section 3 we introduce the matrix
representation of these equations. In sections 4 and 5 we sort out
the toroidal-poloidal components and spherical harmonic
constituents of the matrices. In sections 6 and 7 we give an
ordering of the various components in powers of $\Omega^2$ and
sort out the equations of motions in various orders of magnitude.
References and bibliographical notes relating to mode calculations
in rotating stars are collected in section 7.4. In section 8 we
discuss the numerical results.  A rotating neutron star can be
slowed down by gravitational radiation through  the mass and
mass-current multipole moments of the modes. The modes, in turn,
can be damped out by bulk and shear viscosities present in the
star. The time scales of relevant damping mechanisms are analyzed
in section 9. Calculations of matrix elements and presentation of
appropriate basis sets are given in the
appendices.\\


\section{Review of rotating fluids}\label{rev-rot}

\subsection{Equilibrium configuration}\label{equilib}

Let $\rho(r,\theta)$, $p(r,\theta)$ and $U(r,\theta)$ be the
density, the pressure and the gravitational potential of a star
rotating with the constant angular frequency $\Omega$ about the
z-axis.  Equilibrium condition is

\st\be \nab p+\rho \nab
[U-\frac{1}{2}\Omega^2r^2\cos^2\theta]=0\,. \ee
For slow rotations,
one obtains

\st\be\label{expan} \stq \rho= \rho_0(r)+\Omega^2
[\rho_{20}(r)+\rho_{22}(r)P_2(\cos\theta)], \ee
where $P_2(\cos\theta)$ is a Legendre polynomial.  Similar expansions
exists for $p$ and $U$.  For a barotropic structure,
$p(\rho)$, one will have

\stq\be p_0(\rho_0),\,\, {\rm and} \,\,\,
p_{2i}=\frac{dp_0}{d\rho_0} \rho_{2i};  \,\, i=0,2\,. \ee
Furthermore, Poisson's equation will give

\stq\be \nabla^2( U_0,
U_{20}, U_{22})=4\pi G ( \rho_0,\rho_{20}, \rho_{22})\,. \ee
A thorough study of the structure of rotating polytropes is given
as early as (\cite{Cha33}) by Chandrasekhar.  Further numerical
values on $\rho_{20}$ and $\rho_{22}$ may be found in
Chandrasekhar and Lebovitz (\cite{CLe62}).


\subsection{Linear perturbations}\label{rev-per}

Let a mass element of the rotating fluid at position $\x$, be
displaced by an amount $\xxi^s(\x)\exp(i\omega^s t)$, where, for the
moment, $s$ is the collection of all indices that specify the
displacement in question.  This may include its spherical
harmonic specifications, its radial node number, and/or its
poloidal and toroidal nature.   The Eulerian change in $\rho$,
 and $U$ resulting from this displacement will be

\st\bea\label{den-pre-grav}\stq &&\d^s\rho=-\nab\cdot(\rho\xxi^s)\,,\\ \stq
&&\d^sU(\x)=-G\int\d^s\rho(\xp) \mid\x-\xp\mid^{-1}d^3x'\,. \eea
On the assumption that the displacement takes place adiabatically
and the Lagrangian change in pressure is $-(\partial p/\partial
\rho)_{\rm ad}\rho\nab \cdot\xxi$, one obtains

\stq\bea
\d^sp&&=-(\partial p/\partial \rho)_{\rm
ad}\rho\nab\cdot\xxi^s -\nab p\cdot\xxi^s\nonumber\\
&&=(dp/d\rho)~\d^s\rho +
[dp/d\rho-(\partial p/\partial\rho)_{\rm
ad}]\rho\nab\cdot\xxi^s\,, \eea
where $dp/d\rho$ is the
barotropic derivative of the equilibrium structure. The linearized Euler equation governing
the evolution of $\xxi^s$ now becomes

\st\be\label{lineu} {\cal
W}\xxi^s+2i\omega^s\rho\OOmega\times\xxi^s-{\omega^s}^2\rho\xxi^s=0\,,
\ee
where the linear operator ${\cal W}$ is given by

\stq\be {\cal
W}\xxi^s=\nab\d^sp-\frac{1}{\rho}\nab p\d^s\rho+\rho\nab\d^s U\,.
\ee


\section{Matrix form of equations of motion}\label{mat-eq-motion}

Equation (\ref{lineu}) in its integro-differential form is highly
complicated. We convert it into a
set of linear algebraic equations by expanding $\xxi^s$ in terms
of a complete set of known basis vectors, $\{\zzeta^r(\x)\}$.  Thus

\st\be\label{defxi} \xxi^s=\zzeta^rZ^{rs}\,, \ee
where $Z^{rs}$ are the expansion coefficients and, as yet, are unknown.
Associated with this basis, we define the following matrices

\st\bea\label{defmat} \stq
S^{rs}&&=\int\rho~{\zzeta^r}^*\cdot\zzeta^s d^3x\,,\label{defS}\\
\stq
C^{rs}&&=+i/\Omega\int\rho~{\zzeta^r}^*\cdot(\OOmega\times\zzeta^s)d^3x
\nonumber\\ &&=-i\OOmega/\Omega\cdot\int\rho~{\zzeta^r}^*
\times\zzeta^s d^3x\,,\label{defC}\\ \stq
W^{rs}&&=\int{\zzeta^r}^*\cdot{\cal W}\zzeta^s
d^3x\,,\label{defW} \eea where the integration is over the volume
of the star.  All these matrices are Hermitian.
Furthermore, let
$Z=[Z^{rs}]$ be the matrix of the expansion coefficients, and
$\omega=[\omega^s\d^{rs}]$ be the diagonal matrix of the eigenvalues.
Substituting Eq. (\ref{defxi}) in Eq. (\ref{lineu}), multiplying the
resulting equation from left by ${\zzeta^q}^*$, say, and
integrating over the volume of the star gives the ($q\,s$) element
of the following matrix equation \st\be\label{eqmat} WZ+2\Omega C
Z\omega-S Z\omega^2=0\,. \ee It is important to note that all
factors in Eq. (\ref{eqmat}) are matrices and should not be
commuted with each other without due care.  Solutions of Eq.
(\ref{eqmat}) are equivalent to those of Eq. (\ref{lineu}).  The
eigenfrequencies $\omega$ in both equations are the same, and once
the eigenmatrix $Z$ is known, the set of eigendisplacement vectors
$\{\xxi^s\}$ can be constructed by Eq. (\ref{defxi}).


\section{Partitioning into poloidal and toroidal fields}\label{par-pt}

The basis set $\{\zeta^r\}$ can be divided into two poloidal and
toroidal subsets, $\{\zzeta^r_p\mid \zzeta^r_t\}$.  The set of the
eigenmodes, $[\xxi^s]$, of Eq. (\ref{lineu}) in the absence of
rotation also has such exact partitioning.  Its poloidal subset
comprise the commonly known $g$- and $p$- modes of the fluid.  The
toroidal subset of it includes those displacements of the fluid
which do not perturb the equilibrium state of the star. For
them $\omega=0$. For the sake of mathematical completeness one
might say that $\omega=0$ is a degenerate eigenfrequency
and the set of all toroidal motions $[\xxi^s_t]$ are its eigendisplacement
vectors.

In the presence of rotation two things happen. a) Each known
poloidal mode of the fluid acquires a small toroidal component. b)
The neutral toroidal displacements organize themselves, into a new
sequence of modes and the degeneracy of $\omega=0$ gets removed.
Nonetheless, one may still partition the eigensets $[\xxi^s]$ as
$[\xxi^s_p\mid \xxi^s_t]$ remembering that the subsets $[\xxi^s_p]$
and $[\xxi^s_t]$, unlike the no rotation case, are only
predominantly poloidal and toroidal, respectively.  In view of
these considerations, Eq. (\ref{defxi}) partitions as
\st\bea\label{parxi} \stq
&&\xxi^s_p=\zzeta^r_pZ^{rs}_{pp}+\zzeta^r_tZ^{rs}_{tp}\,,\label{parxip}\\
\stq
&&\xxi^s_t=\zzeta^r_pZ^{rs}_{pt}+\zzeta^r_tZ^{rs}_{tt}\,,\label{parxit}
\eea or in its matrix form and suppressing the superscripts one
gets \stq\be\label{parmatxi}
[\xxi_p\mid\xxi_t]=[\zzeta_p\mid\zzeta_t]\ls \begin{array}{cc}
                                            Z_{pp}&Z_{pt}\\
                                            Z_{tp}&Z_{tt}
                                             \end{array}\rs.
\ee Accordingly, all matrices in Eqs. (\ref{defmat}) partition into
four $pp$, $pt$, $tp$, and $tt$ blocks.  For example, an element
$S^{rs}_{pt}$ in the $pt$ block is obtained by inserting the two
vector $\zzeta^r_p$ and $\zzeta^s_t$ in Eq. (\ref{defS}) and
carrying out the integration.

 We are not interested in the
poloidal modes of Eq. (\ref{parxip}).  They are discussed in ample
details and in a much wider scope than that of the present work
in Sobouti (\cite{Sob80}).  Here we concentrate on the
toroidal modes of Eq. (\ref{parxit}).  The required information
comes from multiplying the block partitioned forms of all the
matrices in Eq. (\ref{eqmat}) and extracting the $tt$ and $pt$
blocks of it.  Thus \st\bea\label{block} \stq
&&tt\rm{-block \hspace{.13cm} of \hspace{.13cm} Eq.  (\ref{eqmat}):}\nonumber\\ &&\;\;W_{tt}Z_{tt}+2\Omega
C_{tt}Z_{tt}\omega_t-S_{tt}Z_{tt}\omega_t^2\nonumber\\
&&\;\;+W_{tp}Z_{pt}+2\Omega
C_{tp}Z_{pt}\omega_t-S_{tp}Z_{pt}\omega_t^2=0\,, \label{ttblock}\\
\stq &&pt\rm{-block \hspace{.13cm} of \hspace{.13cm} Eq.  (\ref{eqmat}):}\nonumber\\ &&\;\;W_{pt}Z_{tt}+2\Omega
C_{pt}Z_{tt}\omega_t-S_{pt}Z_{tt}\omega_t^2\nonumber\\
&&\;\;+W_{pp}Z_{pt}+2\Omega
C_{pp}Z_{pt}\omega_t-S_{pp}Z_{pt}\omega_t^2=0\,. \label{ptblock}
\eea


\section{Partitioning by spherical harmonic numbers}\label{par-sph}

For a toroidal basis vector we will adopt the following spherical
harmonic form \st\bea\label{defzeta} \stq
\zzeta^{r\ell}_t&&=\nab\times
[\hat{r}\phi^{r\ell}(r)Y_{\ell}^m(\vartheta,\varphi)] \nonumber\\
&&=\frac{\phi^{r\ell}}{r}(
0,\frac{im}{\sin\vartheta}Y_{\ell}^m, -\frac{\partial
Y_{\ell}^m}{\partial \vartheta})\,.\label{zeta-t} \eea For a
poloidal vector we will take \stq\be\label{zeta-p}
\zzeta^{r\ell}_p
=\frac{1}{r}[\psi^{r\ell}(r)Y_{\ell}^m,\chi^{r\ell}(r)
\frac{\partial Y_{\ell}^m}{\partial \vartheta},
\chi^{r\ell}(r)\frac{im}{\sin\vartheta}Y_{\ell}^m
]\,. \ee
Appropriate ansatz for the radial function $\phi$, $\psi$, and $\chi$ are given
in appendix B.
The toroidal vector (\ref{zeta-t}) is obviously derived
from a radial vector potential.  The poloidal vector
(\ref{zeta-p}) is actually the sum of two vectors, one derived
from a scalar potential and the other derived from a toroidal
vector potential. See Sobouti (\cite{Sob81}).

In Eqs.
(\ref{defzeta}) we have suppressed the harmonic index $m$ from $\zzeta$'s.  For,
a slowly rotating star is axially symmetric.  Vectors with different
values of $m$ are not mutually coupled.   Vectors belonging to the
same $m$, but different $\ell$'s, however, are coupled.   This
feature entails a further partitioning of the basis sets into
their harmonic subsets,
$[\zzeta^{\ell}_p,\, \ell=0,1,2,\cdots\,]$ and $[\zzeta^{\ell}_t,\,
\ell=1,2,\cdots\,]$.  Correspondingly, each of the matrices in Eqs.
(\ref{block}) partitions into blocks, designated by a pair of
harmonic numbers $(k,\ell)$, say.  For example, the $rs$ element of
$S^{k\ell}_{pt}$, say, will be obtained from Eq. (\ref{defS}) by
inserting the two vectors ${\zzeta^{rk}_p}^*$ and
${\zzeta^{s\ell}_t}$ in that equation.   In the following we
rewrite Eqs. (\ref{block}) taking into account the new
partitioning.  Thus,\\
$tt$-block:
\st\bea\label{hblock} \stq
&&\sum_{\ell'}[~ W^{k\ell'}_{tt}Z^{\ell'\ell}_{tt} +2\Omega
C^{k\ell'}_{tt}Z^{\ell'\ell}_{tt}\omega^{\ell}_t
-S^{k\ell'}_{tt}Z^{\ell'\ell}_{tt}{\omega^{\ell}_t}^2\nonumber\\
&&+W^{k\ell'}_{tp}Z^{\ell'\ell}_{pt} +2\Omega
C^{k\ell'}_{tp}Z^{\ell'\ell}_{pt}\omega^{\ell}_t
-S^{k\ell'}_{tp}Z^{\ell'\ell}_{pt}{\omega^{\ell}_t}^2~]=0,
\label{httblock}\eea
$pt$-block:
\stq\bea &&\sum_{\ell'}[~ W^{k\ell'}_{pt}Z^{\ell'\ell}_{tt}
+2\Omega C^{k\ell'}_{pt}Z^{\ell'\ell}_{tt}\omega^{\ell}_t
-S^{k\ell'}_{pt}Z^{\ell'\ell}_{tt}{\omega^{\ell}_t}^2\nonumber\\
&&+W^{k\ell'}_{pp}Z^{\ell'\ell}_{pt} +2\Omega
C^{k\ell'}_{pp}Z^{\ell'\ell}_{pt}\omega^{\ell}_t
-S^{k\ell'}_{pp}Z^{\ell'\ell}_{pt}{\omega^{\ell}_t}^2~]=0.
\label{hptblock} \eea We again emphasis that each of the factors
in Eqs. (\ref{hblock}) are matrices in their own right.


\section{Expansion order and
spherical harmonic structure of the various
matrices}\label{order-matr}

Expansions of $\rho$, $p$, and $U$ in powers of $\Omega^2$ results
in a corresponding expansion of all the matrices in Eqs.
(\ref{hblock}). Moreover, having the spherical harmonics forms of
Eqs. (\ref{defzeta}), integrations over $\vartheta$ and $\varphi$
dependencies in the calculation of matrix elements can be performed
analytically.  These two tasks are carried out in  appendix A.
The results are quoted below:
\st\bea\label{Stt}
&&\hspace{-.5cm}S^{k\ell}_{tt}=S^{\ell\ell}_{0tt}\d_{k\ell}
+\Omega^2S^{k\ell}_{2tt}
(\d_{k\ell}+\d_{k,\ell\pm 2}),~{\rm see}~  
(\ref{S0tt-int}-\ref{S2tt-int}),\\
\st\label{Cmat} \stq\label{Ctt}
&&\hspace{-.5cm}C^{k\ell}_{tt} =C^{\ell\ell}_{0tt}\d_{k\ell}
+\Omega^2C^{k\ell}_{2tt}
(\d_{k\ell}+\d_{k,\ell\pm 2}),{\rm see}(\ref{C0tt-int}-\ref{C2tt-int})\\
\stq\label{C0tt}
&&\hspace{-.5cm}C^{\ell\ell}_{0tt}=(m/\ell(\ell+1))S^{\ell\ell}_{0tt}\,, 
\hspace{1.95cm}{\rm see}  (\ref{C0tt-int}),\\
\stq\label{Ctp}
&&\hspace{-.5cm}C^{k\ell}_{tp}={C^{\ell k}_{pt}}^*
=C^{k\ell}_{0tp}\d_{k\ell\pm 1}+
{\cal O}(\Omega^2)
\,, \hspace{.45cm}{\rm see}  (\ref{C0pt-int}),\\
\st\label{Wmat}\stq\label{Wpp}
&&\hspace{-.5cm}W^{k\ell}_{pp}=W^{\ell\ell}_{0pp}\d_{k\ell}+{\cal
O}(\Omega^2)\,, \hspace{1.8cm}{\rm see}  (\ref{W0pp-int}),\\
\stq\label{Wtp}
&&\hspace{-.5cm}W^{k\ell}_{tp}={W^{\ell k}_{pt}}^*
=\Omega^2 W^{k\ell}_{2tp}\d_{k\ell\pm 1}
\,, \hspace{1.15cm}{\rm see}  (\ref{W2pt-int}),\\
\stq\label{Wtt}
&&\hspace{-.5cm}W^{k\ell}_{tt} =\Omega^4 W^{k\ell}_{4tt}\,
(\d_{k\ell}+\d_{k,\ell\pm 2})\,,\hspace{1.2cm}{\rm see}  (\ref{W4tt-int}).
\eea
A subscript, 0, 2, 4 preceding the
$tt$, $tp$, $pt$ or $pp$ designations of the matrices indicates the
order of $\Omega$ in the matrix in question. We also note that as
$\Omega\rightarrow 0$, $\omega_t$ does the same. Therefore, it must
have the following form \st\be\label{omega-t}
\omega^\ell_t=\Omega(\omega^\ell_{0t}+\Omega^2
\omega^\ell_{2t})\,. \ee Substituting these order of magnitude
informations in Eq. (\ref{hptblock}) reveals that
\st\be\label{Zpt-tt} \stq\label{Zpt}
Z^{k\ell}_{pt}=\Omega^2Z^{k\ell}_{2pt}\,. \ee For
$Z^{k\ell}_{tt}$, we have no information so far.  Therefore, we
assume the general form \stq\be\label{Ztt}
Z^{k\ell}_{tt}=Z^{k\ell}_{0tt}+\Omega^2Z^{k\ell}_{2tt}\,. \ee


\section{Expansion of equations of motion}\label{order-eq-motion}

Equations (\ref{Stt})-(\ref{Zpt-tt}) allow a consistent expansion
of Eqs. (\ref{hblock}) at $\Omega^2$ and $\Omega^4$ orders and
enable one to decipher the information contained in them.


\subsection{$\Omega^2$ order of the $tt$-block}\label{2nd-tt}

At $\Omega^2$ order Eq. (\ref{httblock}) gives \st\be\label{tt2}
2C^{kk}_{0tt}Z^{k\ell}_{0tt}
-S^{kk}_{0tt}Z^{k\ell}_{0tt}\omega^\ell_{0t}=0\,. \ee For
$k=\ell$, considering the proportionality of $C^{\ell\ell}_{0tt}$
and $S^{\ell\ell}_{0tt}$, Eq. (\ref{C0tt}), one obtains
\st\bea\label{zeromode} \stq\label{omega-0}
\omega^\ell_{0t} &&=2m/\ell(\ell+1)  I,\,\,\, I=\rm{unit \hspace{.13cm} matrix}\,,\\
\stq\label{Z0tt-ll-1}
Z^{\ell\ell}_{0tt} &&\rm{undetermined \hspace{.13cm} at \hspace{.13cm} this \hspace{.13cm} stage}.
\eea
For $k\neq\ell$, using
Eq. (\ref{C0tt}) and (\ref{omega-0}), one has
\stq\be\label{Z0tt-kl}
2m[1/k(k+1)-1/\ell(\ell+1)]
S^{kk}_{0tt}Z^{k\ell}_{0tt} =0\,\, \rm{or} \hspace{.16cm}  Z^{k\ell}_{0tt}=0\,.
\ee Let us summarize the findings so far from a pedagogical point of view.
At $\Omega^2$ order one
solves the eigenvalue Eq. (\ref{tt2}).  In this equation the
Coriolis forces remove the degeneracy of the neutral motions and
create a sequence of modes of purely toroidal nature.  The new
modes have a definite $\ell$-symmetry, (Eq. \ref{zeromode}).  They
are not coupled with toroidal motions of $k\neq\ell$ symmetry (that is
$Z^{k\ell}_{0tt}=0$) and
with poloidal motion (that is $Z^{k\ell}_{0pt}=0$, see Eq. 16a).
 Removal of degeneracy, however, is partial. For, of the
three designations of a standing wave in three dimensions, only
$(\ell,m)$ designations have appeared in the expression for
$\omega^\ell_{0t}=2m/\ell(\ell+1)$.  A third designation,
indicating the radial wave number, is as yet absent.  They will
appear at higher orders of $\Omega$.

One simplifying
feature: We note that $\omega^\ell_{0t}$ of Eq. (\ref{omega-0}) is
a constant matrix.  Therefore, it will commute with any matrix
carrying the same $\ell\ell$ designations, such as
$Z^{\ell\ell}_{0tt}$, $S^{\ell\ell}_{0tt}$, etc.  This feature
was  used in the derivation of Eq.(18c) and will be used repeatedly
in what follows to
simplify the matrix manipulations.


\subsection{$\Omega^2$ order of $pt$-block}\label{2nd-pt}

Equation (\ref{hptblock}) at order $\Omega^2$ along with Eqs.
(\ref{zeromode}) gives \st\be\label{pt2} \stq\label{pt2eq}
W^{kk}_{0pp}Z^{k\ell}_{2pt}+(W^{k\ell}_{2pt}+
2C^{k\ell}_{0pt}\omega^\ell_{0t})Z^{\ell\ell}_{0tt}=0, k=\ell\pm
1\,. \ee $W^{kk}_{0pp}$ is associated with poloidal modes of the
non rotating fluids. In fact if $\zzeta_p$'s are chosen to be the
exact eigenvectors of the non rotating star, $W^{kk}_{pp}$ will be
a diagonal matrix whose elements are the square of the
eigenfrequencies of the $p$- and $g$- modes.  At least in the case
of $p$-modes, $W^{kk}_{0pp}$ is invertible (see Sobouti
\cite{Sob80} for complications in the case of $g$-modes).  Thus,
one obtains \stq\be\label{Z2pt-def}
Z^{k\ell}_{2pt}=-{(W^{kk}_{0pp})}^{-1}[ W^{k\ell}_{2pt}+
C^{k\ell}_{0pt}\omega^\ell_{0t}] Z^{\ell\ell}_{0tt},   k=\ell\pm
1. \ee 
 By Eq.(8b), Eq. (19b) expresses that a toroidal mode of the
rotating fluid of $\ell$ symmetry acquires a small poloidal
component of $\ell\pm 1$ symmetry at $\Omega^2$ order.

 The case of $g$- modes is different. Rotation no matter how small
cannot be treated as a perturbation on them. For, they are created
by minute buoyancy forces and the low frequency tail of their
spectrum gets completely wiped out by any other force in the
medium, here the Coriolis forces. This, in mathematical language
means that $W_{0pp}$ for $g$- modes, is not invertible and Eq.
(19b) is not applicable to them. The way out of the dilemma is to
consider the sum of buoyancy and other intruding forces as an
inseparable entity, without dividing it to large and small
components. The works of Provost et al. (1980) and of Sobouti
(1977, 1980) are examples of such treatments. We leave it to the
experts in the field to decide whether the scrutiny of $g$- modes
in rotating neutron stars is a crucial or an irrelevant issue.\\


\subsection{$\Omega^4$ order of $tt$-block}\label{4th-tt}

\subsubsection{The $\ell\ell$-subblock}\label{4th-tt-ll}

A systematic extraction of $\Omega^4$ order terms of the
$\ell\ell$-block of Eq. (\ref{httblock}) with the help of Eqs.
(\ref{Stt})-(\ref{Zpt-tt}) and elimination of $Z^{k\ell}_{2pt}$ term
appearing in it by Eq. (\ref{Z2pt-def}) gives \st\be\label{tt4}
T^{\ell\ell}_{4tt}Z^{\ell\ell}_{0tt}-{2m\over
\ell(\ell+1)}S^{\ell\ell}_{0tt}
Z^{\ell\ell}_{0tt}\omega^\ell_{2t}=0\,, \ee where the fourth order
$T$-matrix is \stq\bea\label{def-T4tt} T^{\ell\ell}_{4tt}&&=[
W_{4tt}+(2C_{2tt}-S_{2tt}\omega_{0t})\omega_{0t}
\nonumber\\ &&\hspace{-1cm}
-(W_{2tp}+2C_{0tp}\omega_{0t})W^{-1}_{0pp}
(W_{2pt}+2C_{0pt}\omega_{0t})]^{\ell\ell}. \eea Equation
(\ref{tt4}) is a simple eigenvalue problem. Vanishing of its
characteristic determinant, \stq\be\label{tt4-b} \mid
T^{\ell\ell}_{4tt}-{2m\over\ell(\ell+1)}S^{\ell\ell}_{0tt}
\omega^\ell_{2t}\mid=0\,, \ee will give the non degenerate second
order eigenvalues $\omega^{s\ell}_{2t}$, $s=$ radial wave number.
Once they are known, Eq. (\ref{tt4}) itself can be solved for the
eigenmatrix $Z^{\ell\ell}_{0tt}$.  We note that we have solved two
eigenvalue problems, Eqs. (\ref{tt2}) and (\ref{tt4}), to remove
all degeneracies of the zero frequency toroidal motions of the non
rotating fluid.
Further extension of the
analysis of equations of motion to orders higher than $\Omega^4$
will result is non-homogeneous algebraic equations whose
non-homogeneous terms are given in terms of the matrices calculated
in the previous orders.


\subsubsection{The ($\ell, \ell\pm 2$)-subblock}\label{4th-tt-kl}

The presence of $\d_{k,\ell\pm 2}$ in Eqs. (\ref{Ctt}) and
(\ref{Wtt}) indicates that two toroidal motions belonging to
$\ell$ and $\ell\pm 2$ are mutually coupled.  Likewise, the
presence of $\d_{k,\ell\pm 1}$ in Eqs. (\ref{Ctp}) and (\ref{Wtp})
shows the coupling of toroidal and poloidal motion of $\ell$ and $\ell\pm
1$ symmetry.  This brings in an additional
coupling between two toroidal motions of $\ell$ and $\ell\pm 2$
symmetries through the intermediary of poloidal motions.
Therefore, the only unexplored blocks of
Eq. (\ref{httblock}) are those with ($\ell,\ell\pm 2$)
designations. As in section \ref{4th-tt-ll} above, we extract the
$\Omega^4$ order terms of Eq. (\ref{httblock}), but this time with
$\ell$, $\ell\pm 2$ superscripts, eliminate $Z^{\ell',\ell'\pm
1}_{2pt}$ appearing in it by Eq. (\ref{Z2pt-def}) and arrive at

\st\be\label{Z2tt-1}
T^{\ell,\ell\pm 2}_{4tt}Z^{\ell\pm 2,\ell\pm
2}_{0tt}=[ 2C^{\ell\ell}_{0tt}Z^{\ell,\ell\pm 2}_{2tt}-
S^{\ell\ell}_{0tt}Z^{\ell,\ell\pm 2}_{2tt}\omega^{\ell\pm
2}_{0t}] \omega^{\ell\pm 2}_{0t}\,, \ee
where
\stq\bea\label{def-Tlk}
T^{\ell,\ell\pm 2}_{4tt}&&=[
W_{4tt}+(2C_{2tt}-S_{2tt}\omega_{0t})\omega_{0t}
\nonumber\\
&&\hspace{-1cm}
-(W_{2tp}+{4m\over(\ell\pm 2)(\ell\pm 2+1)}C_{0tp})~W^{-1}_{0pp}\nonumber\\
&&~~(W_{2pt}+{4m\over(\ell\pm 2)(\ell\pm 2+1)}C_{0pt})]^{\ell,\ell\pm 2}\,.
\eea

In deriving this expression in two occasions we have substituted for
$\omega^{\ell\pm 2}_{0t}$ and shifted the scalar factor
$2m/(\ell\pm 2)(\ell\pm 2+1)$ across the other matrices.
Returning
to Eq. (\ref{Z2tt-1}) we substitute for $C^{\ell\ell}_{0tt}$ from Eq.
(\ref{C0tt}) and solve for $Z_{2tt}$.  Thus,
\st\be\label{def-Z2tt}
Z^{\ell,\ell\pm
2}_{2tt}=[(\omega^\ell_{0t}-\omega^{\ell\pm 2}_{0t})
\omega^{\ell\pm 2}_{0t}]^{-1}
(S^{\ell\ell}_{0tt})^{-1}T^{\ell,\ell\pm 2} Z^{\ell\pm 2,\ell\pm
2}_{0tt}\,. \ee
Equations (\ref{tt4}), (\ref{Z2tt-1}) and (\ref{def-Z2tt}) are all the
informations contained in the $\Omega^4$ order of the $tt$-block.
Whether $Z^{\ell,\ell}_{2tt}$ is non zero or
otherwise is not clear at this level.  To answer the question one
has to go to $\Omega^4$ and $\Omega^6$ orders of Eqs.
(\ref{hptblock}) and (\ref{httblock}), respectively. This,
however, will not be attempted here.


\subsection{Bibliographical notes}

Papaloizou and Pringle (\cite{PPr78}) have studied the low
frequency $g$- and $r$- modes of Eq. (\ref{lineu}) in an
equipotential coordinate system with their applicability to the
short period oscillations of cataclysmic variables in mind.

Sobouti (\cite{Sob80}) has studied the problem primarily with the goal of
analyzing the perturbative effects of slow rotations on
$p$-modes and demonstrating that rotation, no matter how small,
cannot be treated as a perturbation on $g$-modes.  He argues that
the $g$-modes are fragile structures and their low frequency tail
of the spectrum, below the rotation frequency of the star, will be
completely wiped out by Coriolis and asphericity forces of the
star.  The criterion for the validity of perturbation
expansion is that
the perturbing operator should be smaller than the initial unperturbed
operator everywhere in the Hilbert space spanned by the
eigenfunctions of the unperturbed operator, (Rellich, \cite{Rel69}).
This condition is not met
by $g$-modes when exposed to rotation, magnetic, tidal forces, etc.
For, they have vanishingly small eigenfrequencies.

Provost et al. (\cite{Pro81}) present an analysis of what they call
the ``quasi toroidal modes of slowly rotating stars ''. Their work
should be noted for the consistency of mathematical manipulations
exercised throughout the paper.  They noted
that in neutrally convective rotating stars one
cannot have modes with predominantly toroidal motions. They get mixed
with the neutral convective displacements.

Lockitch
and Friedman (\cite{LFr99}) also address the hybrid modes with
comparable toroidal and poloidal motions.  Their work should be
noted for the emphasis put on the ($\ell,\,m$) parities of the
hybrid components that get coupled through asphericity forces.

Yoshida and Lee (\cite{YLe00}), study Eq. (\ref{lineu}) for a)
those modes that are predominantly toroidal in their nature and b)
for those that have comparable poloidal and toroidal components.
Their latter modes are the same as those of Provost et al. (\cite{Pro81}).



\section{Numerical result for rotating polytropes}

The bulk properties of some observed neutron stars seem to
approximate those of a polytrope of index 1. See Sterigioulas
(1998). It has become fashionable to categories neutron stars as
stiff or soft ones depending on whether their density
distributions are similar to those of polytropes of index smaller
than 1.5 or larger, respectively. To have an example of each
category, sample calculations are given for polytropes of indices
1 and 2.

The structure of rotating polytropes, taken from Chandrasekhar
(\cite{Cha33}), is summarized in appendix C. The ansatz for the
scalars $\phi^{s\ell}$, $\psi^{s\ell}$ and $\chi^{s\ell}$
appearing in Eqs. (\ref{defzeta}) are given in appendix B.  The
required matrix elements are reduced in appendix A.  For a given
$N=1,2,\ldots $, the $N\times N$ matrices are numerically
integrated. The eigenvalues $\omega^\ell_{0t}$ and
$\omega^\ell_{2t}$ are calculated from Eqs. (\ref{omega-0}) and
(\ref{tt4-b}).  Once the eigenvalues are known the various
components of the eigenmatrices $Z^{\ell\ell}_{0tt}$, $Z^{\ell\pm
2,\ell}_{2tt}$ and $Z^{\ell\pm 1,\ell}_{2pt}$ are calculated from
Eqs. (\ref{tt4}), (\ref{def-Z2tt})and (\ref{Z2pt-def}). The
results are given in tables 1 to 3.

In table 1, to show the convergence of the variational
calculations, the eigenvalues $\omega^{\ell}_{2t}$ are displayed
for polytrope of index 2, $\ell=m=2$, and for $N=1,2,3,4,5$. This
table should be considered as a basis for judging the accuracy of
the numerical values. With only five variational parameters, the
first, second and third eigenvalues are produced with an accuracy
of few parts in $10^{5}$, $10^{4}$ and $10^{2}$, respectively.
Likewise, the numerical values in the remaining tables should be
trusted to the same degree of accuracy and for the first few
modes.

In table 2, the second order eigenvalues, $\omega^{\ell}_{2t}$ and
coefficient matrices, $Z^{\ell\ell}_{0tt}$,
$Z^{\ell+1,\ell}_{2pt}$, and $Z^{\ell+2,\ell}_{2tt}$ are given for
polytrope of index 1 and for $\ell=m=2, N=5$. In table 3 we give the same
calculations for polytrope of index 2. The eigenvalues are in
units of $\sqrt{\pi G {\bar\rho}}$. They are in agreement with
those of Lindblom et al. (\cite{LMO99}), Yoshida and Lee
(\cite{YLe00}), and Morsink (\cite{Mor01}). Each of these authors
have used their own technique, and different from that of the
present paper.

A novel feature of the present analysis is the provision of much
detail on eigendisplacement vectors, information that can be
profitably used to calculate any other bulk or local parameter of
the model. For example, for modes belonging to $\ell=m=2$ one may
write 
\st\bea\label{xi-2} 
\xxi^{s,2}_t=\sum_r\zzeta^{r,
2}_t(Z^{2,2}_{0tt})^{rs}+\Omega^2 \sum_r\zzeta^{r,
3}_p(Z^{3,2}_{2pt})^{rs}~~~~~~\nonumber\\
+\Omega^2 \sum_r\zzeta^{r,
4}_t(Z^{4,2}_{2tt})^{rs}\,, \eea 
in which the first sum is the
backbone of the mode and is of zero order. It is toroidal motion
of $\ell=m=2$ symmetry.  The second sum is the coupling of
$\ell=3,\,m=2$ poloidal motion with $\ell=m=2$ toroidal motion. It
is a poloidal motion and is of second order. The third sum is the
coupling of $\ell=4,\,m=2$ toroidal motion with $\ell=m=2$
toroidal motion and again is of second order. In figures 1 to 4 we
have plotted the radial behavior of
$\sum_r\zzeta^{r,2}_t(Z^{2,2}_{0tt})^{rs}$, $\sum_r\zzeta^{r,
3}_p(Z^{3,2}_{2pt})^{rs}$, and $\sum_r\zzeta^{r,
4}_t(Z^{4,2}_{2tt})^{rs}$ for polytropes of indices $1,2$ and the
first two modes, $s=1,2$. The center and the surface are nodes in
all curves. For $s=2$ there is an extra node in between in every
curve. The general rule is; number of nodes for any parameter
$f(r, \theta, \varphi)$ belonging to the radial mode number $s$,
is $s+1$. This includes the ever present nodes at the center and
the surface of the star. This feature is faithfully present in all
five modes that can be constructed from the data of table 1 and 2,
even though we know that the numerical values for $s=4$ and 5 are
only orders of magnitude.

\section{r-mode time scales}
In this section we study the dissipative effects of viscosity and
gravitational radiation on r-modes. Quite generally and regardless
of whether the star rotates or not, the total energy of an
undamped mode, $\xxi(r,t)$, is \st\bea E &&= {1\over 2} \int [
\rho\, \dot{\xxi}_{real} \cdot \dot{\xxi}_{real} + \xxi_{real}
\cdot {\cal W} \xxi_{real}] d^3x,
\\ &&=~maximum~kinetic~energy~of~the~mode\,
\nonumber \label{energy} \eea
See appendix D for proof of Eq.(24).
In the presence of viscous forces and the gravitational radiation
the combined rate of dissipation (Cutler and Lindblom, 1987) is
\st
\bea {d E\over dt} = -\int(2\eta\delta
\sigma^*_{ab}\delta\sigma^{ab} +\zeta\delta\sigma^*
\delta\sigma)d^3x\hspace{2.5cm}\nonumber\\
-\omega_t(\omega_t-m\Omega)\sum_{\ell\geq
2}N_\ell(\omega_t-m\Omega)^{2\ell} (|\delta D_{\ell m}|^2+|\delta
J_{\ell m}|^2)\,, \label{energy-evol} \eea
where $\eta$ and
$\zeta$ are the shear and bulk viscosities respectively,
\st\stq
\be \delta \sigma_{ab}={\scriptstyle {1\over 2}}
(\nabla_{\!a}\dot\xi_b+\nabla_{\!b}\dot\xi_a -{\scriptstyle {2\over
3}}\delta_{ab}\nab\cdot\dot\xxi), ~~~{\rm shear~strain},
\label{shear-def}
\ee
\stq\bea
\delta\sigma
&&=\nabla\cdot\dot\xxi\,~~~{\rm bulk~strain},\label{bulk-def} \\
N_\ell
&&= {4\pi G\over c^{2\ell+1}} {(\ell+1)(\ell+2)\over
\ell(\ell-1)[(2\ell+1)!!]^2}, ~~~~c=~{\rm speed~of~light},\,
\nonumber \label{12}
\eea
\st\be
\d D_{\ell m}=\int\d\rho\, r^\ell
Y^*_{\ell\,m} d^3x, ~~~{\rm mass~multipole~moment},
\label{mass-def}
\ee
\st\bea
\d J_{\ell m}= {2\over
c}\sqrt{\ell/(\ell+1)} \int (\rho \,\dot\xxi +\d\rho\, \v)\cdot
\Y^*_{\ell\,m} d^3x, ~~~\nonumber\\{\rm current~multipole~ moment},
\label{current-def}
\eea
\stq\bea
\Y_{\ell\,m}={1\over\sqrt{\ell(\ell+1)}} \nabla \times (\hat{\r}
Y_{\ell\,m}),~~~~~~~~~~ 
\nonumber\\~~~~{\rm toroidal~vector~spherical~harmonics},
\eea
For a mode of the type of Eq.(23) the bulk viscous dissipation
$(bv)$ comes entirely from its poloidal component, as
$\nab\cdot\zzeta_t=0$. For $m=\ell$ this has the following rate
\st\be
\label{bulk-evol} (d E^{s\ell}/dt)_{bv}= -\omega^{\ell
2}_{0t} \Omega^6 \{Z^{k\ell\dagger}_{2pt}(bv)
 Z^{k\ell}_{2pt} \}^{ss}, k=\ell+1
\ee 
where the elements of the new matrix, $(bv)$, are
\stq\bea\label{bulk-e} 
&&\hspace{-.4cm}(bv)^{rq}= \int
\zeta[d(r\psi^{rk})/dr-k(k+1)\chi^{rk}]\nonumber\\  
&&\hspace{2cm} [ d(r\psi^{qk})/dr-k(k+1)\chi^{qk}]dr/r^3. 
\eea
The shear viscous
dissipation has contributions from both toroidal and poloidal
components of Eq.(23). To the lowest order in $\Omega$, however,
the toroidal component has the dominant contribution. Thus,
\st\be\label{shear-evol} (d E^{s\ell}/dt)_{sv}= -\omega^{\ell
2}_{0t}\Omega^2 \{ Z^{\ell\ell\dagger}_{0tt} (sv)
Z^{\ell\ell}_{0tt}\}^{ss}, \ee where 
\stq\bea\label{shear-o}
(sv)^{rq} = \ell(\ell+1) \int
\eta[(r{d\phi^{r\ell}/dr}-2\phi^{r\ell})
(r{d\phi^{q\ell}/dr}-2\phi^{q\ell})\nonumber\\ +
(\ell^2+\ell-2)\phi^{r\ell}\phi^{q\ell} {dr/r^2}].~~~~~~~ \eea 
Dissipation
due to the gravitational radiation, to the lowest order, comes
from the current multipole moment, $\delta^r J_\ell$. This has the
following vector elements 
\st\be\label{shear-ov}
\d^rJ_\ell=i\omega^\ell_{0t}\Omega{2\ell\over c} \int
\rho_0\,r^{\ell+1} \phi^{r\ell} dr. \label{current1} \ee To obtain
the dissipation time scales: a)we have calculated the integrals of
Eqs.(24)-(31), numerically. b) For the bulk and shear viscosities
we have adopted the values of Cutler and Lindblom, $\eta = 347
\rho^{9/4} T^{-2}{\rm g~cm^{-1}~s^{-1}},~~ \zeta=6.0\times
10^{-59} \rho^2 \omega^{-2}_t T^6 {\rm g~cm^{-1}~s^{-1}},$ where
$T$ is the temperature. For $a$ mode of radial node number $s$ and
$\ell=m=2$, we obtain 
\st\bea\label{time}
{1\over\tau^s(\Omega,T)}={1\over{\tilde\tau}^s_{sv}}\lp{10^9
K\over T}\rp^2 +{1\over{\tilde\tau}^s_{bv}}\lp{T\over 10^9
K}\rp^6\lp{\Omega^2\over\pi G{\bar\rho}}\rp
\nonumber\\
+{1\over{\tilde\tau}^s_{gr}}\lp{\Omega^2\over \pi
G{\bar\rho}}\rp^3,~~~~~ \eea 
where $\bar\rho$ is the average density of
the star.  Here ${\tilde\tau}^s_{sv}$, ${\tilde\tau}^s_{bv}$, and
${\tilde\tau}^s_{gr}$ are the shear viscous-, bulk viscous- and
the gravitational radiation- time scale, respectively, which are
normalized for $T=10^9$ K and $\Omega^2=\pi G\bar\rho$. The total
dissipation time scale is \st\be\label{time1}
{1\over\tau(\Omega,T)}=\sum_s {1\over\tau^s(\Omega,T)}. \ee For a
model of $1.4 M_\odot$ and $R=12.53 {\rm km}$ we have calculated
${\tilde\tau}^s_{sv}$, ${\tilde\tau}^s_{bv}$, and
${\tilde\tau}^s_{gr}$ for $s=1,2,3,4,5$ and displayed in the unit
of time in tables 4 and 5.  Our values for $s=1$ are in agreement
with those of Lindblom et al. (\cite{LMO99}) and, Yoshida and Lee
(\cite{YLe00}). To the best of our knowledge, the values for
$s\geq 2$ are new.

In a newly born hot neutron star, $T\geq 10^{12}K$ the bulk
viscosity has a dominant role in damping out the perturbations and
cooling down the star. In colder stars, $T\leq 10^{10}$ K and
$\Omega^2\sim \pi G \bar\rho$, the gravitational radiation is more
important than radiation shear and bulk counterparts. While
driving the $r$-modes unstable, it spins down the star. It is
believed that the star looses much of its energy and angular
momentum through the gravitational radiation in this stage. In a
case study of Andersson and Kokkotas (2000), the rotational period
increases from 2 ms to 19 ms in one year. Below $T=10^8$ K and
$\Omega^2\sim \pi G {\bar\rho}/100 $, the shear viscosity is the
dominant factor in cooling down the star.


\begin{acknowledgements}

We wish to thank Sharon Morsink for a careful reading of the
manuscript and illuminating comments.  We are pleased to 
acknowledge the referee for his valuable comments. 

\end{acknowledgements}


\appendix


\section{The elements of $S, C$ and $W$ matrices}\label{app-A}

In calculating the elements of various matrices the following parameters
and
integrals are encountered frequently: \st\bea
&&Q_\ell=[(\ell^2-m^2)/(2\ell-1)(2\ell+1)]^{1/2}\,,
\eea
\st\bea &&\int\cos\vartheta  Y^*_{km}Y_{\ell
m}d\Omega=Q_{\ell+1}\d_{k,\ell+1}
+Q_\ell \d_{k,\ell-1}\,,\nonumber\\  
&& ~~~~~~~~~~d\Omega=\sin\vartheta d\vartheta d\varphi\,,
\eea
\st\bea
&&\int\sin\vartheta  Y^*_{km}\partial Y_{\ell m}/\partial\vartheta
d\Omega
=\ell Q_{\ell+1}\d_{k,\ell+1}-(\ell+1)Q_\ell\d_{k,\ell-1}\,,\nonumber\\
\eea
\st\bea
&&\int[(m^2/\sin^2\vartheta)  Y^*_{km}Y_{\ell m}+
\partial Y^*_{km}/\partial\vartheta
\partial Y_{\ell m}/\partial\vartheta]\cos\vartheta d\Omega\nonumber\\
&&  =\ell(\ell+2)Q_{\ell+1}\d_{k,\ell+1}+(\ell^2-1)Q_\ell\d_{k,\ell-1}\,,
\eea
\st\bea
&&\int\cos^2\vartheta  Y^*_{km}Y_{\ell m}d\Omega\nonumber\\
&&  =(Q^2_{\ell+1}+Q^2_\ell)\d_{k\ell}
+Q_{\ell-1}Q_\ell\d_{k,\ell-2}
+Q_{\ell+1}Q_{\ell+2}\d_{k,\ell+2}\,,\nonumber\\
\eea
\st\bea
&&\int[(m^2/\sin^2\vartheta) Y^*_{km}Y_{\ell m}+
\partial Y^*_{km}/\partial\vartheta
\partial Y_{\ell m}/\partial\vartheta] P_2(\cos\vartheta)d\Omega\nonumber\\
&&  ={3\over 2}[\ell(\ell+3)Q^2_{\ell+1}+(\ell-2)(\ell+1)Q^2_\ell
-{1\over 3}\ell(\ell+1)]\d_{k\ell}\nonumber\\
&&  +{3\over 2}[(\ell-2)(\ell+1)Q_{\ell-1}Q_\ell \d_{k,\ell-2}+
\ell(\ell+3)Q_{\ell+1}Q_{\ell+2} \d_{k,\ell+2}]\,.\nonumber\\
\eea

The basic definitions for $S$, $C$, and $W$ matrices are given in
Eqs. (\ref{defmat}).  Expansions of $\rho$, $p$, and $U$ entering
these defining integrals are in Eqs. (\ref{expan}).  The zero and
$\Omega^2$ orders of these variables are sufficient to carry out
the analysis of section \ref{order-eq-motion} up to $\Omega^4$
order consistently.

Finally, the spherical harmonics forms of the
basis toroidal and poloidal vectors are given in Eqs.
(\ref{defzeta}).  Angular integrals entering the definition of any
matrix element at any desired order are performed analytically.
Integrals in radial directions are left for numerical
calculations.\\

\noindent The $S$-matrix:
\st\bea\label{S0tt-int}
(S^{k\ell}_{0tt})^{rs}=\d_{k\ell} \ell(\ell+1)\int^R_0\rho_0
\phi^{r\ell}\phi^{s\ell}
dr\,, \eea
\st\bea\label{S2tt-int}
&&\hspace{-.4cm}(S^{k\ell}_{2tt})^{rs}=\d_{k\ell} \ell(\ell+1)
\int^R_0\rho_{20}\phi^{r\ell}\phi^{s\ell}dr\nonumber   \\
&&\hspace{-.2cm}+{3\over 2}\{\d_{k\ell}
[\ell(\ell+3)Q^2_{\ell+1}+(\ell-2)(\ell+1)Q^2_\ell
-{1\over 3}\ell(\ell+1)]\nonumber\\
&&\hspace{-.2cm}+\d_{k,\ell-2} (\ell-2)(\ell+1)Q_{\ell-1}Q_\ell
+\d_{k,\ell+2} \ell(\ell+3)Q_{\ell+1}Q_{\ell+2}\}\nonumber\\
&&\hspace{2cm}\times\int^R_0\rho_{22}\phi^{rk}\phi^{s\ell}dr\,.
\eea
The $C$-matrix:
\bea\label{C0tt-int}
&&\hspace{-.4cm}(C^{k\ell}_{0tt})^{rs}=
(m/\ell(\ell+1))(S^{k\ell}_{0tt})^{rs}\,,
\eea
\st\bea\label{C2tt-int}
&&\hspace{-.4cm}(C^{k\ell}_{2tt})^{rs}=\d_{k\ell} m
\int^R_0\rho_{20}\phi^{r\ell}\phi^{s\ell}dr\nonumber\\ 
&&\hspace{-.2cm}+{9\over
2}m[\d_{k\ell}  (Q^2_{\ell+1}+Q^2_\ell -1/9)
+\d_{k,\ell-2} Q_{\ell-1}Q_\ell\nonumber\\
&&\hspace{.9cm}+\d_{k,\ell+2} Q_{\ell+1}Q_{\ell+2}]
\int^R_0\rho_{22}\phi^{rk}\phi^{s\ell}dr\,.\\
\stq\label{C0pt-int}
&&\hspace{-.4cm}(C^{k\ell}_{0pt})^{rs}={(C^{\ell k}_{0tp})^{rs}}^*\nonumber\\
&&\hspace{.9cm}=-i \d_{k,\ell+1} \ell  Q_{\ell+1}
\int^R_0\rho_0[\psi^{rk}+(\ell+2)\chi^{rk}]\phi^{s\ell}dr\nonumber\\
&&\hspace{1cm}+i \d_{k,\ell-1} (\ell+1) Q_\ell
\int^R_0\rho_0[\psi^{rk}-(\ell-1)\chi^{rk}]\phi^{s\ell}dr\,.\nonumber\\
\eea 
The $W$-matrix: 
\st\bea\label{W0pp-int}
&&\hspace{-.4cm}(W^{k\ell}_{0pp})^{rs}=
\d_{k\ell} \{ \int^R_0 \rho^{-1}_0 dp_0/
d\rho_0 \d^{r\ell}_p\rho_0(r) \d^{s\ell}_p\rho_0(r)  dr/r^2
\nonumber\\ 
&&+\int^R_0[(\partial p_0/\partial\rho_0)_{\rm ad}- d
p_0/d\rho_0]\rho_0(d\psi^{rk}/dr-\chi^{rk}/r)\nonumber\\
&&\hspace{2cm}\times~
(d\psi^{s\ell}/dr-\chi^{s\ell}/r) dr/r^2\}\,, 
\eea
\st\bea\label{W2pt-int} 
&&\hspace{-.4cm}(W^{k\ell}_{2pt})^{rs}={(W^{\ell
k}_{2tp})^{rk}}^*\nonumber\\ 
&&\hspace{.9cm}=-3im[ \d_{k,\ell+1}
Q_{\ell+1}+\d_{k,\ell-1} Q_\ell]\nonumber\\
&&\hspace{1.2cm}\times \int^R_0 \rho^{-1}_0 dp_0/
d\rho_0 \rho_{22}\d^{rk}_p\rho_0(r) \phi^{s\ell} dr/ r^2\,, 
\eea
where 
\st\bea 
&&\hspace{-.4cm}\d^{r\ell}_p\rho_0(r)=[(1/\rho_0) (d/dr)
(r\rho_0\psi^{r\ell})- \ell(\ell+1)\chi^{r\ell}]\,,\nonumber 
\eea
\st\bea\label{W4tt-int} 
&&\hspace{-.4cm}(W^{k\ell}_{4tt})^{rs}=9m^2[\d_{k\ell} (
Q^2_{\ell+1}+Q^2_\ell) +\d_{k,\ell-2} Q_{\ell-1}Q_\ell\nonumber\\
&&\hspace{-.3cm}+\d_{k,\ell+2} Q_{\ell+1}Q_{\ell+2}
]\int^R_0 \rho^{-1}_0 dp_0/d\rho_0\rho^2_{22}
\phi^{rk}\phi^{s\ell} dr/r^2.
\eea 
Poloidal motions give rise to
density perturbations and therefore to perturbations in self
gravitation.  In Eqs. (\ref{W0pp-int}) and (\ref{W2pt-int}) the
latter is neglected (Cowling's approximation) for simplicity.
Otherwise there is no conceptual difficulty in
including another term in Eqs. (\ref{W0pp-int}-13)  
to avoid this approximation.

Finally, we note that the upper limit of all radial integrals here
is the radius of non rotating star instead of that of the rotating
one.  We reproduce the argument of Sobouti (\cite{Sob80}) to show
that, in most cases, the effect arising from this difference in the limits of
integrations is of far higher order in $\Omega^2$ than to upset the
analysis of this paper. Let $\Delta R(\vartheta)=R(\vartheta)-R$
be the distance between two points with coordinate $\vartheta$ and
situated on the surfaces of rotating and non rotating stars.
Obviously $\Delta R$ is of $\Omega^2$ order.  In a typical error
integral, $\int_R^{R+\Delta R}f(r) dr$, we expand $f(r)$ about $R$
and obtain
\st\be\label{error} \int_R^{R+\Delta R}f(r)
dr=f(R)\Delta R +{1\over 2}f'(R) (\Delta R)^2+\cdots\,. \ee
In
Eqs. (\ref{S0tt-int}-14) 
the integrands depend on a combination of the variables $\rho_0$,
$\rho_{20}$, $\rho_{22}$, $p_0$ and $p_{22}$.  For a star of
effective polytropic index $n$, at the surface $\rho_{0}$ and
$p_{0}$ vanish as $(R-r)^n$ and $(R-r)^{n+1}$. The leading error
terms of Eq. (A.15) are $(\triangle R)^{n+1}\propto
\Omega^{2(n+1)}$ and $(\triangle R)^{n+2}\propto \Omega^{2(n+2)}$,
respectively. The distorted quantities $\rho_{20}, \rho_{22}$ and
$p_{20}, p_{22}$ tend to zero as $(R-r)^{n+1}$ and $(R-r)^{n}$.
However, considering the fact that these are second order
quantities and should be multiplied by an extra factor of
$\Omega^{2}$ wherever they appear, the leading error term in
expressions involving them are also of the order $\Omega^{2(n+1)}$
and $\Omega^{2(n+2)}$. Thus, the largest error committed in
replacing the volume of the rotationally distorted star by that of
the non rotating one is of the order $\Omega^{2(n+1)}$, a matter
of no concern for the analysis of this paper if $n\geq1$.


\section{Ansatz for the scalars $\phi^{s\ell}$, $\psi^{s\ell}$ and
$\chi^{s\ell}$}

The vicinity of the center of a star is a uniform medium, in the
sense that, as $r$ tends to zero, $\rho(r)$, $p(r)$,
$U(r)$, etc. all tend to constant values.  Any scalar function,
$\sigma(r)$
say, associated with a wave in such a nondispersive uniform and
isotropic environment should satisfy the wave equation $\nabla^2
\sigma(r)+k^2\sigma(r)=0$, $k=$const. Furthermore, if this scalar
is associated with the spherical harmonic $\ell$, ie. if it is
of form $\sigma(r)Y_\ell^m(\vartheta,\varphi)$ and is finite at
the origin, should tend to zero as $r^\ell$.  Therefore $\sigma$
should have an expansion of the form
$\sigma(r)=r^\ell\sum_{s=0}^\infty a_s r^{2s}$.  This is how the
solutions of Laplace's equation ($k=0$), spherical Bessel function
and many other hypergeometric functions behave.  A spherical harmonic vector, $\xxi^\ell$
belonging to $\ell$, quite generally can be written in terms of
three scalars
\st\be \xxi^\ell=-\nab(\sigma_1Y_\ell^m)+\nab\times
{\bf A},     {\bf
A}=\hat{r}\sigma_2Y_\ell^m+\nab\times(\r\sigma_3Y_\ell^m)\,, \ee
where $\sigma_i,\,i=1,2,3$ are scalars of the type described above.
Therefore, the radial and non radial components of $\xxi$
should have the form
\st\be \xi_r, \xi_\vartheta, \xi_\varphi
\rightarrow r^\ell \sum_{s=1} b_s r^{2s-1}\,. \ee
To ensure this
behavior, it is sufficient that the scalars $\phi^{r\ell}$,
$\psi^{r\ell}$, and $\chi^{r\ell}$ entering Eqs. (\ref{defzeta})
to be proportional to $r^{\ell+2r}$.

We adopt
the following ansatz for the
\st\be\label{B1}
(\phi^{r\ell}, \psi^{r\ell},
\chi^{r\ell})=\theta(r) r^{\ell+2 r},   r=0,1,2,\ldots
\ee
where $\theta(r)$ is the polytropic function and was found by trial and error that
ensures faster convergence of the variational calculations.
The ansatz has
the required $r^\ell$ behavior at the center.
Two remarks are in order here:

1) That a power set $\{r^{\ell+2r}, r=0,1,2,\ldots\}$ is complete
for expanding any function of $r$ that behaves as $r^\ell$ near the
origin follows form a theorem of Weiresstraus
(Relich \cite{Rel69}, Dixit et al. \cite{Dix}).

2) We have chosen the ansatz of Eq. (\ref{B1}) for their
simplicity. They are not the most efficient ones for rapid
convergence of variational calculations.  The set of the
asymptotic expressions that helioseismogists use for
eigendisplacement vectors in the sun and other stars would,
perhaps, give a faster convergence of the numerical computations,
see Christensen-Dalsgaard (\cite{Dal97}) and references therein.


\section{Review of rotating polytropes}

The structure of rotating polytropes is taken from a landmark
paper of Chandrasekhar (\cite{Cha33}).  A summary of what is
needed here with slight changes in his notation is as follows
\st\bea\label{C1} \stq\label{C1a} \rho_0 &&=\rho_c\theta^n,
\rho_{20}=n\rho_c\theta^{n-1}\Psi_0,
\rho_{22}=n\rho_c\theta^{n-1}\Psi_2\,, \\
\stq\label{C1b}
p_0 &&=p_c\theta^{n+1},  p_{20}=(n+1)p_c\theta^n\Psi_0,
p_{22}=(n+1)p_c\theta^n\Psi_2\,,\nonumber\\
\eea
where $\rho_c$
and $p_c$ are constants, $\theta$ is the polytropic variable and
satisfies the Lane-Emden equation \st\be\label{C2}
(1/\eta^2)(d/d\eta)(\eta^2d\theta/d\eta)
=-\theta^n,    \eta=[(n+1)p_c/4\pi G\rho_c^2]r\,, \ee
and $\Psi_0$, $\Psi_2$ satisfy the following \st\bea\label{C3}
\stq\label{C3a}
(1/\eta^2)(d/d\eta)(\eta^2d\Psi_0/d\eta)
&&=-n\theta^{n-1} \Psi_0+1\,,\\
\stq\label{C3b}
(1/\eta^2)(d/d\eta)(\eta^2d\Psi_2/d\eta) &&=(
-n\theta^{n-1} +6/\eta^2)\Psi_2\,.
\eea
It should be noted
that Chandrasekhar's rotation parameter is $\Omega^2/2\pi
G\rho_c$.  For the purpose of this paper we have integrated Eqs.
(\ref{C2}-5) 
numerically.


\section{The energy of normal mode}
Let us take the real part of Eq. (4) and write it in the following form
\st\be\label{D1}
{\cal W} \xxi_{\rm re} + 2 \rho \dot{\xxi}_{\rm re} \times \OOmega
+\rho \ddot{\xxi}_{\rm re} =0
\ee
We take the scalar product of Eq. (\ref{D1}) by $\dot{\xxi}_{\rm re}$ and
integrate
over the volume of the star.
The Coriolis term gives no contribution. 
The first term considering the hermitian
character of ${\cal W}$ gives
\st\bea\label{D2}
\int \dot{\xxi}_{\rm re} \cdot {\cal W} \xxi_{\rm re}d^3x
&& ={1\over 2} \int (\dot{\xxi}_{\rm re} \cdot {\cal W} \xxi_{\rm re}
+\xxi_{\rm re} \cdot {\cal W} \dot{\xxi}_{\rm re})d^3x \nonumber\\
&& ={1\over 2} {d \over dt} \int \xxi_{\rm re} \cdot {\cal W}
\xxi_{\rm re} d^3x. 
\eea 
The third term in Eq. (\ref{D1}) gives
\st\be\label{D3} 
\int \rho \dot{\xxi}_{\rm re} \cdot
\ddot{\xxi}_{\rm re} d^3x = {1 \over 2}{d \over dt} \int \rho
\dot{\xxi}_{\rm re} \cdot \dot{\xxi}_{\rm re} d^3x. 
\ee 
Equations (\ref{D1})
and (\ref{D2}) should add to zero, which after a time integration gives
the constant total energy 
\st\be\label{D4} 
E=E_{\rm kin} +E_{\rm
pot}={1\over 2} \int (\rho \dot{\xxi}_{\rm re} \cdot
\dot{\xxi}_{\rm re} +\xxi_{\rm re} \cdot {\cal W} \xxi_{\rm re})
d^3x. 
\ee 
Next we substitute for ${\cal W}\xxi_{\rm re}$ from
Eq. (\ref{D1}). After simple manipulations, we obtain 
\st\bea\label{D5} 
E && = {1 \over 2} \int \rho (\dot{\xxi_{\rm re}}\cdot
\dot{\xxi_{\rm re}}- \xxi_{\rm re} \cdot\ddot{\xxi_{\rm re}})
d^3x\nonumber\\
&&=
{1\over 2} \int \rho [2 \dot\xxi_{\rm re}\cdot \dot\xxi_{\rm re} -
{\partial \over \partial t}
(\xxi_{\rm re}\cdot \dot\xxi_{\rm re})] d^3x\nonumber\\
&& = 2 E_{\rm kin} - {1 \over 4} {d^2 \over dt^2} \int \rho \xxi_{\rm re}
\cdot \xxi_{\rm re} d^3x. 
\eea
Since the time dependence of $\xxi_{\rm re}$ is sinusoidal, 
upon taking the time average
of Eq. (\ref{D5}) the second integral vanishes and are obtains
\st\be\label{D6}
E=2\overline{E_{\rm kin}} = E_{\rm kin}({\rm maximum}), \hspace{1cm}
{\rm QED}.
\ee


\newpage

\noindent {\bf Figure Captions:}\\
\begin{itemize}
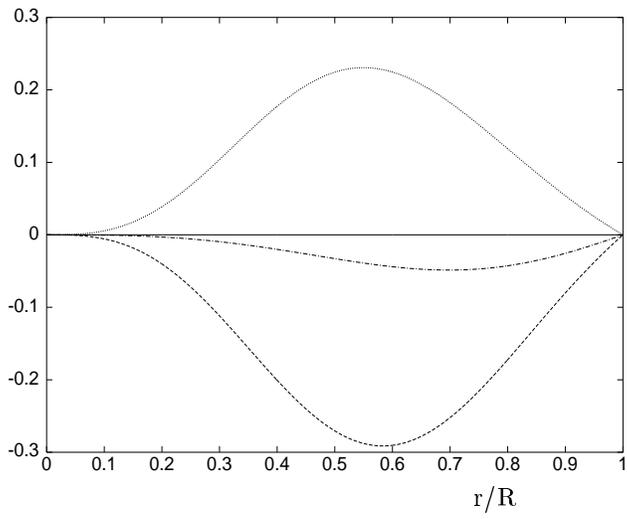
\item{Fig. 1:} Radial behavior of the various components of the eigenfunction
of Eq.(23) for $\ell=m=2, s=1, n=1. \sum_r \zzeta^{r,2}_t(Z^{2,2}_{0tt})^{rs}$,
dashed curve; $\sum_r \zzeta^{r,3}_p(Z^{3,2}_{0pt})^{rs}$, dot-dashed curve;
$\sum_r \zzeta^{r,4}_t(Z^{4,2}_{2pt})^{rs}$, dotted curve. Data for $Z$'s are taken from the
first column of Table 2. Nodes in all three components are at the center and surface.

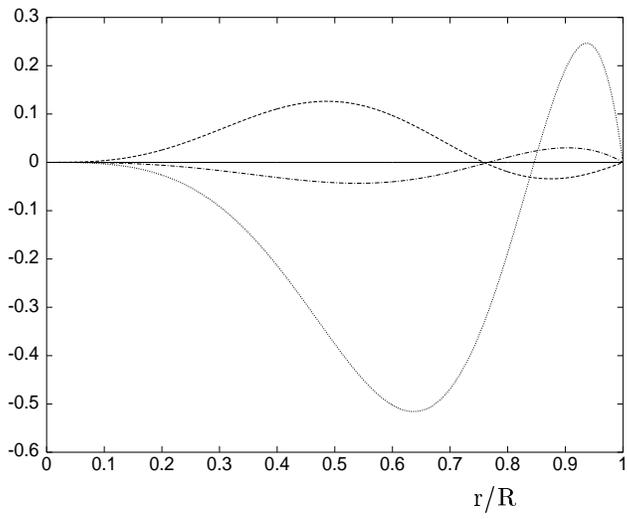
\item{Fig. 2:} Same as Fig. 1 for $\ell=m=2, s=2, n=1$. Data for $Z$'s
are from the second column of Table 2. Note the extra node in all three components.

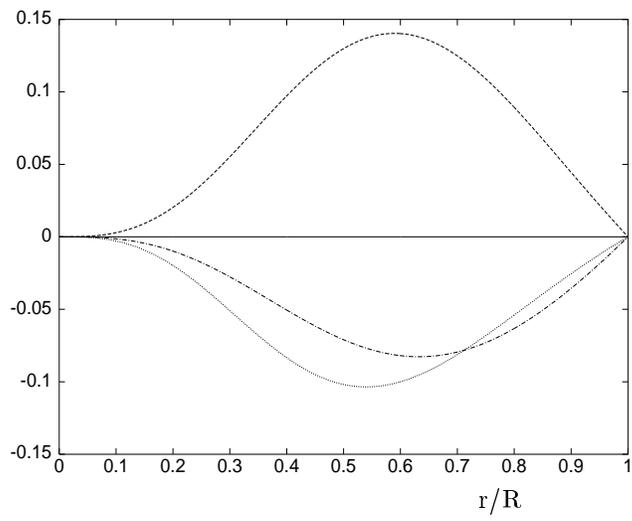
\item{Fig. 3:} Same as Fig. 1 for $\ell=m=2, s=1, n=2$. Data for $Z$'s
are from the first column of Table 3. Nodes are at the center and the surface.

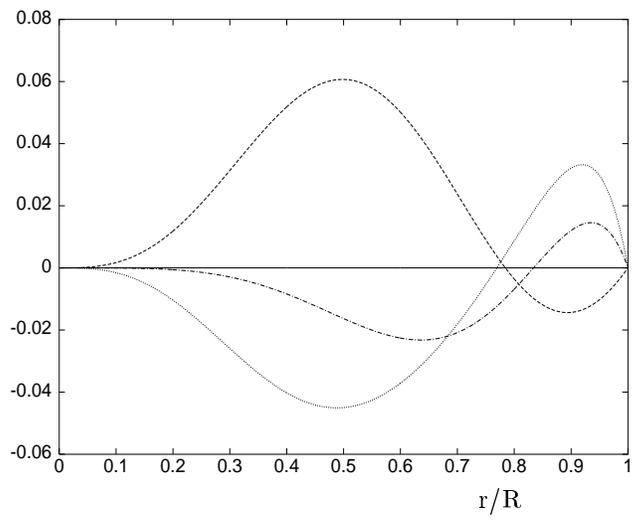
\item{Fig. 4:} Same as Fig. 1 for $\ell=m=2, s=2, n=2$. Data for $Z$'s
are from the second column of Table 3. Note the extra nodes
in all components.\vspace{1cm}\\

\end{itemize}

\noindent {\bf Table Captions:}\\
\begin{itemize}
\item{Table 1:} Convergence of the variational calculations. For polytrope of
index 2 and $\ell=m=2$, the second order toroidal eigenvalues, $\omega^\ell_{2t}$, are
displayed for different number of variational parameters, $N=1, 2, 3, 4, 5$.
They are in units of $\sqrt{\pi G \bar\rho}$. A number $a\times 10^{\pm b}$
is written as $a\pm b$.

\item{Table 2:} Second order eigenvalues $\omega_{2t}$ and coefficient matrices,
$Z^{2,2}_{0tt}$, $Z^{3,2}_{2pt}$, and $Z^{4,2}_{2tt}$ are given for $m=2, N=5$ and
the polytropic index $1$.

\item{Table 3:} Same as table 2, for polytrope 2.

\item{Table 4:} Shear viscous-, bulk viscous-, and gravitational
radiation- time scales in seconds are given for polytrope 1 and
$s=1, 2, 3, 4, 5$ from top to bottom. A number $a\times 10^{\pm
b}$ is written as $a\pm b$.

\item{Table 5:} Same as table 4, for polytrope 2.\vspace{1cm}\\
\end{itemize}

\begin{center}
Table 1\vspace{.52cm}\\
\begin{tabular}{ccccc}\hline
 .30169+0 &&&&\\
 .19860+0 & .60797+0&&&\\
 .16137+0 & .44344+0 & .1371+1 &&\\
 .16136+0 & .43726+0 & .11361+1 & .30568+1&\\
 .16132+0 & .43714+0 & .10106+1 & .24613+1 & .52712+1\\\hline
\end{tabular}\vspace{1cm}\\
 \end{center}

\newpage
\begin{center}
Table 2\vspace{.52cm}\\
\begin{tabular}{cccccc}\hline
 $\omega_{2t}$ & .32203+0 & .84087+0 &  .19790+1 &  .48110+1 & .96363+1\\
$Z_{0tt}^{2,2}$ &&&&&\\
 &-.58565+1 & .38532+1 &-.29986+1 &-.27754+1 & .14743+2\\
 & .12510+2 &-.11134+2 & .82347+1 & .72343+1 &-.73216+2\\
 &-.12234+2 & .84713+1 &-.34999+1 & .18385+0 & .14699+3\\
 & .64179+1 &-.87049+0 &-.73634+2 &-.15379+2 &-.13539+3\\
 &-.15465+1 &-.78018+0 & .49263+1 & .11366+2 & .47334+2\\
$Z_{2pt}^{3,2}$&&&&&\\
&-.39677+0 &-.82952+0 & .96949-1 &-.12394+0 &-.27257+0\\
&-.20041+0 & .98613+0 &-.28392+0 & .65146+0 & .10959+1\\
& .63624+0 & .96043+0 &-.27949+0 &-.14499+1 &-.17471+1\\
&-.41126+0 &-.37431+0 & .42798+0 & .10202+1 & .98804+0\\
& .92562-1 &-.53391-1 &-.13166+0 &-.23806+0 &-.18318+0\\
$Z_{2tt}^{4,2}$&&&&&\\
& .58000+1 &-.30934+1 & .27352+1 & .30368+1 &-.40984+1\\
&-.16462+2 &-.10961+2 &-.10022+2 &-.11591+2 & .18955+2\\
& .21461+2 & .15392+2 & .14924+2 & .18085+2 &-.35467+2\\
&-.14090+2 & .10291+2 &-.10129+2 &-.12810+2 & .30614+2\\
& .37608+1 &-.27390+1 & .26484+1 & .33165+1 &-.10091+2\\\hline
& $s=1$    & $s=2$    & $s=3$    & $s=4$    & $s=5$
\end{tabular}\vspace{2cm}\\
\end{center}

\begin{center}
Table 3\vspace{.52cm}\\
\begin{tabular}{cccccc}\hline
$\omega_{2t}$&.16132+0 & .43714+0 & .10106+1 & .24613+1 & .52712+1\\
$Z_{0tt}^{2,2}$&&&&&\\
& .30158+1 & .17867+1 & .11589+1 & .82841+0 & .10498+2\\
&-.36757+1 &-.25829+1 &-.20167+0 & .15548+1 &-.45626+2\\
& .16613+1 &-.30333+1 &-.85360+1 &-.14096+2 & .84738+2\\
&-.37105+0 & .52330+1 & .14253+2 & .24110+2 &-.75082+2\\
& .14113+0 &-.18708+1 &-.59865+1 &-.13019+2 & .25924+2\\
$Z_{2pt}^{3,2}$&&&&&\\
&-.14392+1 &-.28790-1 & .29383+0 & .19977-1 &-.24544+0\\
& .10270+1 &-.14184+1 &-.11345+1 & .15290+0 & .92290+0\\
&-.28306+0 & .15325+1 & .57987+0 &-.95505+0 &-.14964+1\\
&-.17515-1 & .71869+0 &-.63417-1 & .68145+0 & .73244+0\\
& .20617-1 & .13678+0 &-.15707-1 &-.15094+0 &-.11366+0\\
$Z_{2tt}^{4,2}$&&&&&\\
&-.30726+1 &-
.16563+1 &-.14505+1 &-.14874+1 &-.21505+1\\
& .71922+1 & .48149+1 & .43706+1 & .46605+1 & .81784+1\\
&-.83777+1 &-.58904+1 &-.56669+1 &-.63205+1 &-.13491+2\\
& .51958+1 & .36393+1 & .34767+1 & .40067+1 & .10782+2\\
&-.13588+1 & .93612+0 &-.85413+0 &-.92167+0 &-.34032+1\\\hline
& $s=1$    & $s=2$    & $s=3$    & $s=4$    & $s=5$
\end{tabular}
\end{center}

\newpage
\clearpage
\newpage

\begin{center}
Table 4\vspace{.52cm}\\
\begin{tabular}{cccc}\hline
$s$ & $\tilde\tau_{sv}$ & $\tilde\tau_{bv}$ & $\tilde\tau_{gr}$ \\\hline
1 & 2.14+8 & 1.20+11  &- 2.79+0 \\
2 & 7.78+8 & 1.12+11  &- 8.71-1\\
3 & 1.41+9 & 1.04+11  &- 5.53-1 \\
4 & 1.72+9 & 9.72+10  &- 5.66-1\\
5 & 1.47+8 & 9.06+10  &- 1.00+0 \\\hline
\end{tabular}\vspace{1cm}\\
\end{center}

\begin{center}
Table 5\vspace{.52cm}\\
\begin{tabular}{cccc}\hline
$s$ & $\tilde\tau_{sv}$ & $\tilde\tau_{bv}$ & $\tilde\tau_{gr}$
\\\hline
1 & 1.28+9 & 2.07+10  & -3.60+0 \\
2 & 6.89+8 & 2.84+10  & -3.51+0\\
3 & 1.68+8 & 3.60+10  & -7.52+0 \\
4 & 1.74+9 & 4.36+10  & -5.61+0\\
5 & 6.24+7 & 5.04+10  & -6.68+0\\\hline
\end{tabular}
\end{center}

\end{document}